\documentclass[11pt,twoside]{article}
\usepackage{asp2004}
\usepackage{epsf}
\usepackage{psfig}

\begin{document}
\title{Parsec-scale Constraints on the Ionized Interstellar Medium
  with the Terzan~5 Pulsars}
\author{Scott M. Ransom}
\affil{National Radio Astronomy Observatory, 520 Edgemont Rd.,
  Charlottesville, VA 22901}

\begin{abstract}
  Over the past two years, we used a series of GBT observations to
  uncover at least 33 millisecond pulsars in the globular cluster
  Terzan~5 located in the Galactic bulge. We now have 32 timing
  solutions for the pulsars which give us precise positions and
  dispersion measures (DMs) and indicate that the DMs are dominated by
  variations in the integrated electron density along the slightly
  different sight lines towards the pulsars. At a distance of
  $\sim$8.7 kpc, angular separations between the pulsars range from
  0.4\arcsec$-$100\arcsec\ and correspond to projected physical
  separations of 0.01$-$4\,pc, giving us a unique probe into the
  ionized ISM properties on these scales.  Our measurement of the DM
  structure function toward Terzan~5 is not inconsistent with
  Kolmogorov-like electron density fluctuations in the ISM on scales
  ranging from at least 0.2$-$2\,pc.
\end{abstract}

\section{Introduction}
Terzan~5 is a rich globular cluster (GC) located in the Galactic bulge
($l$=3\fdg8, $b$=1\fdg7) at a distance $d$=8.7$\pm$2\,kpc
\citep{clge02}. Deep and repeated observations of Terzan~5 using the
Green Bank Telescope (GBT) at 2\,GHz since mid-2004 have uncovered 30
new millisecond pulsars (MSPs), bringing the total known in the
cluster to 33, by far the most of any GC \citep{rhs+05,hrs+06}.  Many
individual GC MSPs are in interesting exotic binary systems, yet
ensembles of MSPs in the same cluster can provide unique science as
well.  Since the pulsars are at identical (within $\sim$0.1\%) and
known distances, and are separated by small angles on the sky
($\la$1\arcmin$-$2\arcmin), they are unique probes of pulsar
luminosities and masses \citep[e.g.][]{and92}, cluster dynamics
\citep[e.g.][]{phi92}, and ionized gas --- both within the cluster
\citep{fkl+01} and in the intervening interstellar medium
\citep[ISM;][]{and92}.

We have recently determined timing solutions for 32 of the 33 pulsars
in Terzan~5.  These solutions provide (among other things), positions
on the sky (errors are typically $\sim$0.01\arcsec\ in right ascension
and $\sim$0.1\arcsec$-$0.4\arcsec\ in declination; see Figure~1),
dispersion measures (DMs; the integrated number of electrons along the
line-of-sight to the pulsar) with relative errors
$\la$0.01\,pc\,cm$^{-3}$ and absolute errors $\la$0.1\,pc\,cm$^{-3}$,
and the apparent spin period derivative, $\dot P_{obs}$.  The latter
values are usually dominated by the gravitational acceleration of the
cluster itself, such that values of $\dot P_{obs}$ less than or
greater than zero indicate positions behind or in front of the cluster
respectively \citep{phi92}.  In this work, we use the accurate
positions and DMs to compute the DM structure function, $D_{\rm
  DM}(\delta\theta)$, over a range of angles $1\arcsec \la
\delta\theta \la 100\arcsec$, and to constrain the intervening ionized
ISM on linear scales from $10^{17}\la l \la 10^{19}$\,cm.

\begin{figure}[!ht]
\plotone{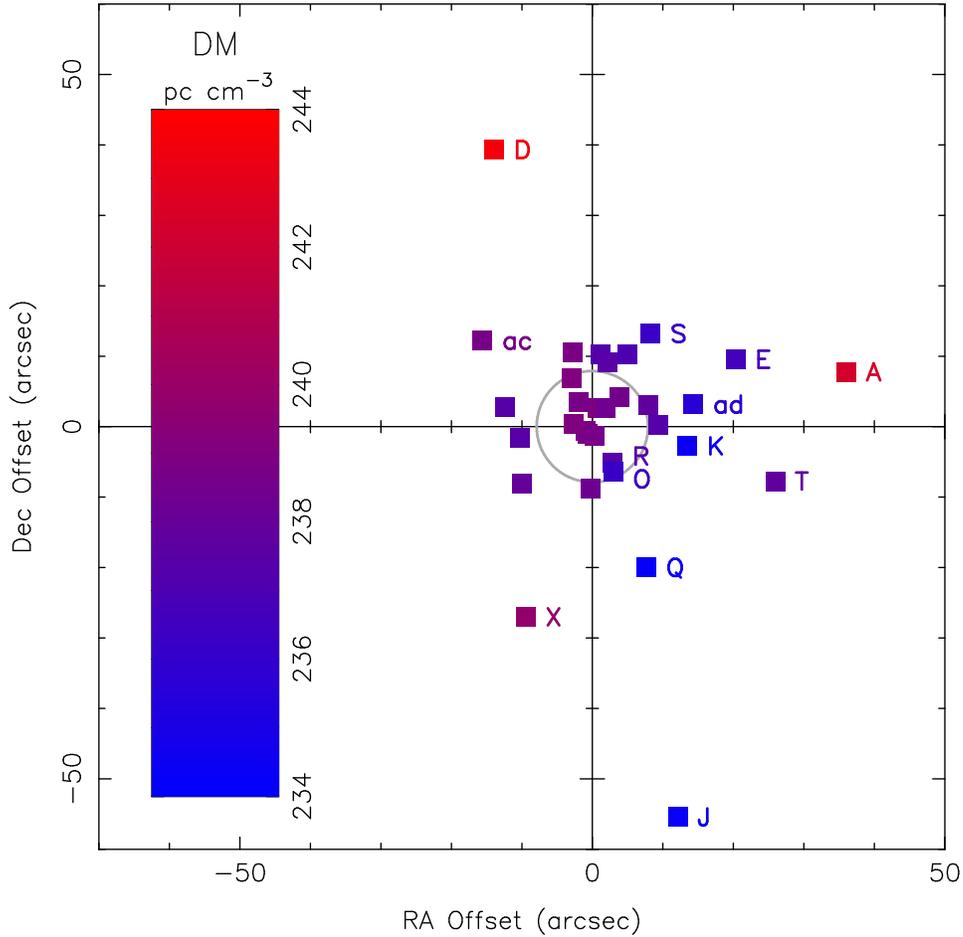}
\caption{Positions and Dispersion Measures (DMs) of the 32 Terzan~5
  pulsars with timing solutions.  Positional errors are typically
  $\sim$0.01\arcsec\ in right ascension, and $\sim$0.1$-$0.4\arcsec\
  in declination.  Relative errors on the DMs are
  $\la$0.01\,pc\,cm$^{-3}$, while the absolute DMs are known to
  $\la$0.1\,pc\,cm$^{-3}$.  The grey circle represents the cluster
  core radius ($r_c$=7.9\arcsec).  The two pulsars Terzan~5~O and
  Terzan~5~R are separated by only 1.2\arcsec\ on the sky, yet their
  DMs are different by 1.1\,pc\,cm$^{-3}$.}
\end{figure}

\section{The DM Structure Function}

Variations in ISM electron density, $\delta n_e$, have been shown to
be consistent with a Kolmogorov-like wavenumber spectrum of the form
$P_{\delta n_e}(q) = C_n^2q^{-\beta}$, with $\beta \sim 11/3$ over a
very wide range of wavenumber $q = 2\pi/l$ \citep*[corresponding to
scales $l$ from 10$^8$$-$10$^{15}$\,cm; e.g.][]{ars95}.  Measurements
of pulsar scintillation, scattering, and DM time variations have been
important throughout this range in $l$.

For simultaneous observations of multiple pulsars separated by small
angles $\delta\theta\ll1$\,rad (i.e. GC MSPs), and assuming isotropic
electron variations, the electromagnetic phase structure function
\citep[e.g.][]{cr98} depends only on the separation of the sources in
the plane of the sky, $\delta r = d\,\delta\theta$, and can be written
$$D_\phi(\delta r) = (r_e\lambda)^2 f_\alpha (d\,\delta\theta)^\alpha 
\int_0^d\left(\frac{s}{d}\right)^\alpha\ C_n^2(s)\ ds,$$ where
$\lambda$ is the radio frequency, $r_e$ is the classical electron
radius, $\alpha=\beta-2$ and $f_\alpha$ is a constant (when
$\alpha$=5/3, $f_\alpha$=88.3).  Following Cordes (private comm), if
we assume that the $C_n^2$ are constant out to distance $d_{\rm eff}$,
and recognize $C_n^2d_{\rm eff}$ as the scattering measure SM($d_{\rm
  eff}$),
$$D_\phi(\delta r) = (r_e\lambda)^2 \frac{f_\alpha}{\alpha+1} 
\left(\frac{d}{d_{\rm eff}}\right)^\alpha {\rm SM}(d_{\rm
  eff})^\alpha.$$

For GC MSPs, we observe DM variations as a function of $\delta\theta$
rather than phase variations.  But we can relate the DM variations to
phase variations using $\delta{\rm DM} = \delta\phi/\lambda r_e$.
Therefore, we effectively measure the phase structure function by
determining the squared DM differences as a function of angular
separation $\delta\theta$, which is the DM structure function:
$$D_{\rm DM}(\delta\theta) = \frac{f_\alpha}{\alpha+1} 
\left(\frac{d}{d_{\rm eff}}\right)^\alpha {\rm SM}(d_{\rm
  eff})^\alpha.$$

For Terzan~5 (or other sources at large distances in the Galactic
plane within the $\sim$1\,kpc scale height of $n_e$), $d \sim d_{\rm
  eff}$.  This means that a good measurement of $D_{\rm
  DM}(\delta\theta)$ over a range of $\delta\theta$ will determine the
power-law index of the electron density variations (from its slope
$\alpha = \beta-2$ on a log-log plot) and an estimate of the
scattering measure and $C_n^2$ (from the $y$-intercept).  To be useful
for ISM constraints, though, measured DM variations must be dominated
by the intervening ISM rather than an ionized intracluster medium as
has been detected in 47~Tucanae \citep{fkl+01}.

As noted recently by \citet{fhn+05}, the total DM spread, $\Delta$DM,
of the pulsars in GCs with large average DMs ($\ga$50\,pc\,cm$^{-3}$)
scales roughly linearly with DM.  If DM variations were dominated by
intracluster gas, $\Delta$DM would be roughly independent of distance
and the average DM.  However, Terzan~5 has one of the largest average
DMs ($\sim$239\,pc\,cm$^{-3}$) and the largest $\Delta$DM (almost
10\,pc\,cm$^{-3}$) of any known cluster.  If intracluster gas was the
cause, its density would be a factor 10$-$20 greater than that found
in 47~Tucanae \citep{fhn+05}.  Another striking indication of the
dominance of the ISM contribution to $\Delta$DM over an intracluster
gas contribution comes from the pulsars Terzan~5~O and Terzan~5~R (see
Figure~1).  These pulsars are separated by only 1.2\arcsec\ on the
sky, yet their DMs are different by 1.1\,pc\,cm$^{-3}$ in the
``wrong'' way to be explained by gas within the cluster.  Terzan~5~O,
which has the smaller DM, is located behind the cluster (as determined
by its negative $\dot P_{obs}$), while Terzan~5~R has the larger DM
and is likely in front of the cluster.  Such a situation can only
occur if the DM variations are dominated by an intervening ISM that is
highly structured on linear scales between 0.01$-$1\,pc.

\begin{figure}[!ht]
\plotone{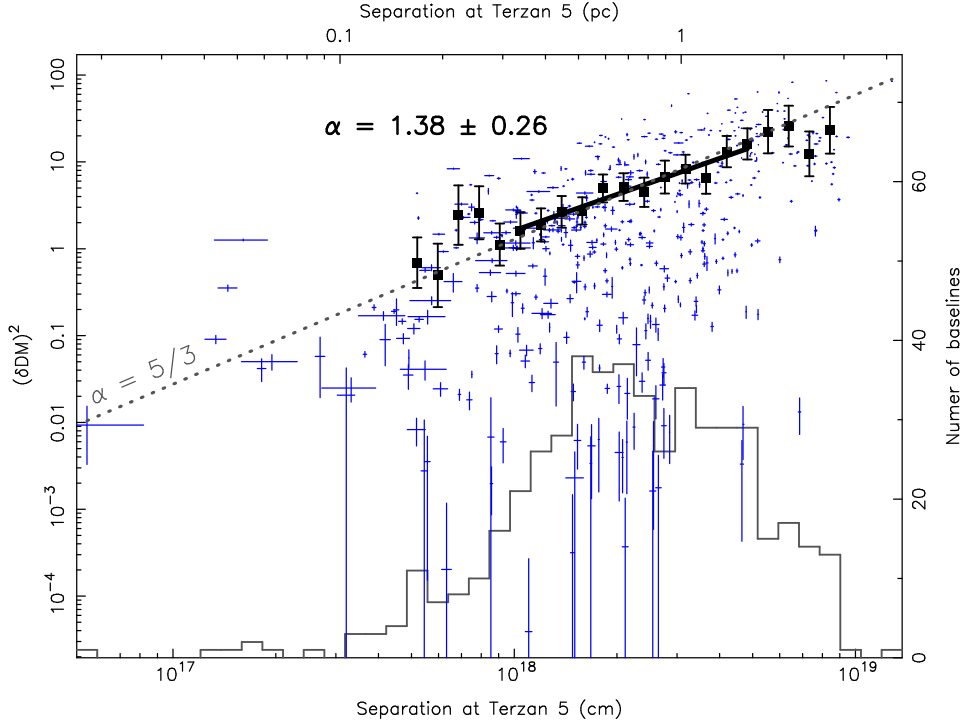}
\caption{The DM structure function, $D_{\rm DM}(\delta\theta)$, for
  the 32 Terzan~5 MSPs with timing solutions.  The 496 measurements of
  $(\delta{\rm DM})^2$ and their errors are the thin crosses.  The
  grey histogram show the number of measurements in 40 evenly spaced
  bins.  For bins with $\ga$6 measurements, the $\left<(\delta{\rm
      DM})^2\right>$ are shown as black boxes with error bars. The
  thick black line shows the linear fit to the bins with $\ga$20
  measurements, which determined the power-law index $\alpha =
  1.38\pm0.26$.  The grey dotted line shows the best fit model
  assuming a Kolmogorov spectrum of electron density variations with
  $\alpha = 5/3$.}
\end{figure}

\section{Data Analysis}

With $n = 32$ independent timing solutions for the MSPs in Terzan~5,
we made $n(n-1)/2 = 496$ measurements of $(\delta{\rm DM})^2$, one for
each pair of MSPs (see Figure~2).  To determine $D_{\rm
  DM}(\delta\theta)$, we binned the full range of separations,
projected to the assumed $d$=8.7\,kpc distance\footnote{Since we are
  probing the intervening ISM, an argument can be made to use a
  fraction of the distance to Terzan~5 rather than the total distance,
  $d$.} to Terzan~5 using $d\,\delta\theta$, into 40 logarithmically
spaced bins.  If there were at least 6 measurements of $(\delta{\rm
  DM})^2$ within a bin, we computed the mean-squared DM difference
$\left<(\delta{\rm DM})^2\right>$.  Errors on $\left<(\delta{\rm
    DM})^2\right>$ were estimated using the measured DM distribution
for Terzan~5 and the number of measurements per bin, according to
normal sampling theory.

For bins with $\ge$20 measurements of $(\delta{\rm DM})^2$, we fit a
simple linear (in log-log space) model to the data:
$\log_{10}\left(\left<(\delta{\rm DM})^2\right>\right) = \alpha
\log_{10}\left(d\,\delta\theta\right) + b$.  For the first model, we
allowed both $\alpha$ and $b$ to vary and measured
$\alpha$=1.38$\pm$0.26.  Since the measured power-law index of the
turbulence is only $\sim$1-$\sigma$ away from the value for Kolmogorov
fluctuations, we determined $b$ by assuming $\alpha$=5/3, and
re-fitting to get $b$=$-$29.893$\pm$0.030.  Applying appropriate unit
conversions, ${\rm SM} = 10^{(b + 29.166)} =
0.188$$\pm$0.013\,kpc\,m$^{-20/3}$, and therefore $C_n^2 = {\rm
  SM}/d_{\rm eff} \sim 0.022$\,m$^{-20/3}$.

\section{Discussion and Conclusions}

It is interesting to compare our estimate of the SM towards Terzan~5,
0.188 $\pm$ 0.013\,kpc\,m$^{-20/3}$, to values determined in other
ways.  \citet{nt92} measured a pulse broadening timescale due to
scattering, $\tau_p$, for pulsar Terzan~5~A at 685\,MHz of 0.70\,ms.
This can be converted to a pulse broadening based SM estimate,
SM$_{\tau}$, using $\tau_p = 0.90\,{\rm ms}\,{\rm
  SM}_\tau^{6/5}\nu^{-22/5}d$, with $\nu$ in GHz and $d$ in kpc
\citep{cor02}, which gives ${\rm SM}_\tau$=0.033\,kpc\,m$^{-20/3}$.
This value is smaller by a factor of $\sim$6 than our measurement,
although we should expect some systematic variation due to the
different weighting factors used by the methods over what is almost
certainly an inhomogeneous electron distribution along the
line-of-sight \citep{cor02}.  It could also be that there is an upturn
in the power spectrum between the small scales probed by $\tau_p$ and
the much longer scales relevant to the DM structure function.

Similarly, we can use either the known $\left<\rm
  DM\right>\sim239$\,pc\,cm$^{-3}$ or distance to Terzan~5,
$d$=8.7\,kpc, as inputs to the NE2001 Galactic electron density model,
which can estimate SM \citep{cl02}.  If we specify $d$, NE2001
predicts SM$_{d}$=0.79\,kpc\,m$^{-20/3}$, while if we specify DM,
SM$_{\rm DM}$=0.29\,kpc\,m$^{-20/3}$.  The latter value is quite close
to our measurement, and significantly closer than the estimate based
on the distance to the cluster\footnote{If the distance to Terzan~5 is
  really 8.7\,kpc, NE2001 greatly overestimates the integrated
  electron density towards the cluster.  For that distance, NE2001
  predicts DM$\sim$530\,pc\,cm$^{-3}$.}.

It is quite interesting that the measured power-law index of the DM
structure function for linear scales ranging from $\sim$0.2$-$2\,pc
(or wavenumbers, $10^{-16}<q<10^{-15}$\,m$^{-1}$) is not inconsistent
with Kolmogorov fluctuations of the ISM.  These scales are
10$^2$$-$10$^3$ times longer than those probed by most other pulsar
techniques.  When we use our estimates of $C_n^2$ and assume $\beta =
11/3$ to determine $P_{\delta n_e}(q)$, the resulting values are
amazingly close to extrapolations based on scintillation and other
measurements at wavenumbers between $10^{-13}<q<10^{-6}$\,m$^{-1}$
\citep{ars95}, implying that Kolmogorov-like density fluctuations are
present in the ISM over size scales covering almost ten decades.  It
is intriguing to note the possible downturn of $D_{\rm
  DM}(\delta\theta)$ at linear scales of 2$-$3\,pc.  While this is
likely due to the poor statistics at the large-angle end of the
structure function, it could also indicate saturation of the structure
function upon reaching the outer scale (i.e. the largest size scale
where power is input into the turbulent process governing the ISM).

Given the large number of pairwise measurements provided by the
Terzan~5 MSPs, one of our next tasks will be to check for evidence of
{\em anisotropic} electron density fluctuations over the same size
scales.  In addition, ongoing searches of GCs using the GBT and other
telescopes have resulted in several other bulge GCs with substantial
numbers of pulsars where these techniques might be useful in the
future (for instance, M28 currently has 11 known pulsars).

\acknowledgements Special thanks go to Jim Cordes, Jean-Pierre
Macquart, Toney Minter, and Avinash Deshpande for helping me to
understand the uniqueness of using cluster MSPs to constrain the
ionized ISM, and for describing how to go about doing just that.  Big
thanks also go to the rest of the Terzan~5 crew (Ingrid Stairs, Jason
Hessels, Paulo Freire, Vicky Kaspi, and Fernando Camilo) for all their
observing, searching, solving, and timing efforts.


\end{document}